\documentclass[aps,prd,eqsecnum,11pt,showpacs,nofootinbib]{revtex4-1}
\usepackage{geometry}                		
\usepackage{graphicx}				
\usepackage{amssymb}
\usepackage{graphicx}
\usepackage{amssymb,amsmath}
\usepackage{amsfonts,mathrsfs}

\newcommand{\be}{\begin{equation}}
\newcommand{\ee}{\end{equation}}
\newcommand{\bea}{\begin{eqnarray}}
\newcommand{\eea}{\end{eqnarray}}
\newcommand{\bes}{\begin{subequations}}
\newcommand{\ees}{\end{subequations}}
\newcommand{\w}{\omega}

\begin{document}

\title{Looking for traces of Hawking radiation\\ in correlation functions of BEC acoustic black holes 
}

\author{Paul~R.~Anderson}
\email{anderson@wfu.edu}
\affiliation{Department of Physics, Wake Forest University, Winston-Salem, North Carolina 27109, USA}
\author{Roberto~Balbinot}
\email{Roberto.Balbinot@bo.infn.it}
\affiliation{Dipartimento di Fisica dell'Universit\`a di Bologna and INFN sezione di Bologna, Via Irnerio 46, 40126 Bologna, Italy}
\author{Richard~A.~Dudley}
\email{Richard.A.Dudley@charlotte.edu}
\affiliation{Department of Physics and Optical Sciences, University of North Carolina at Charlotte, Charlotte, North Carolina 28223, USA}
\author{Alessandro~Fabbri}
\email{afabbri@ific.uv.es}
\affiliation{Departamento de F\'isica Te\'orica and IFIC, Universidad de Valencia-CSIC, C. Dr. Moliner 50, 46100 Burjassot, Spain}


\begin{abstract}

Renaud Parentani was one of the leading figures in Quantum Field Theory in curved spacetime, in particular concerning its applications to Hawking-like radiation in analogue models.  In this paper dedicated to him, we discus the characteristic features appearing in the correlation
functions in an acoustic black hole formed by a Bose-Einstein condensate, considered as signature of the presence of Hawking radiation in this system.


 \end{abstract}

\maketitle

\section{Introduction}

In 1981 Unruh \cite{unruh81} suggested that Hawking's black hole (hereafter BH) radiation  can have an analogue in a fluid whose flow undergoes a transition from a subsonic regime to a supersonic one. The locus where this happens is the so called "sonic
horizon"", since sound waves are trapped inside the supersonic region and cannot propagate upstream; they are trapped by the flow and dragged downstream. Unruh showed that in this situation one should expect an emission in the subsonic
region of thermal phonons at a temperature proportional to the surface gravity of the sonic horizon exactly as predicted by Hawking for gravitational BHs \cite{hawking74, hawking75}.

In the following years many systems were proposed to experimentally detect this analogue Hawking radiation \cite{Barcelo:2005fc}. The most promising appeared to be the ones constructed by Bose-Einstein condensates (BECs), since in this case one can arrange the experimental setup so that
the associated Hawking temperature is expected to be just one order of magnitude smaller than the BEC background temperature ($100\ nK$). Nevertheless even this difference has so far prevented any direct detection of these thermal phonons.

In 2008 it was shown that, 
since the Hawking effect is a genuine process of entangled pair creation  in which for each thermal phonon in the subsonic region there is a corresponding negative energy partner inside the horizon,  a characteristic correlation band in the in-out  density-density correlation function should appear~\cite{paper1, cfrbf}.  This is the smoking gun of the Hawking effect. This band was observed in a series of experiments performed by Steinhauer and his group~\cite{jeff2016, jeff2019, jeff2021}. 
 This represents the best evidence to date for the presence of Hawking radiation in sonic BHs.

Renaud Parentani suggested that besides this main band, two other minor bands should appear in the density-density correlation function because of backscattering effects on the modes \cite{mp}. While there is as yet no experimental evidence for these bands, numerical calculations have confirmed their presence~\cite{paper2, rpc, paper2011}.  

In this paper, dedicated to Renaud Parentani for his invaluable contributions to this field, we shall investigate, using the framework of Quantum Field Theory in curved space (see for example \cite{bd, fu, pt}), the significative features of the density-density correlation function connected to Hawking radiation for two flow profiles having the same asymptotic sound speed limits and horizon surface gravity, outlining similarities and differences.

\section{The setting}

In a BEC the phase fluctuation $\hat\theta$ on top of the condensate in the hydrodynamical approximation obeys an equation which is formally identical to a wave equation for a massless scalar field propagating in a fictitious curved space-time\footnote{More details of the review in this section can be obtained in~\cite{Barcelo:2005fc}.}  
described by the line element
\be \label{dueuno} ds^2=\frac{n}{mc}[-c^2dT^2+(d\vec x - \vec{v} dT)(d\vec x - \vec{v} dT)]\ , \ee
where $n$ is the condensate density, $c$ the local speed of sound, $\vec{v}$ the velocity field and $m$ the mass of a single atom. The wave equation reads
\be\label{duedue}
\Box \hat\theta =
0\ ,
\ee
where $\Box\equiv \nabla_\mu\nabla^\mu$ is the covariant d'Alembertian calculated with the metric (\ref{dueuno}).


This system can be treated by using the methods of Quantum Field Theory in curved space-time. This has been done in a paper written in collaboration with R. Parentani \cite{paper2013}. Here we just outline the main points.

We shall consider for simplicity a stationary unidimensional flow directed along the $\hat x$ axis with a constant velocity $\vec v$, the density  $n$  is also constant.
By an appropriate rescaling of the phase operator  $\hat{\theta}=\sqrt{\frac{mc}{n\hbar L_\perp^2 }}\ {\hat{\theta}^{(2)}}$, where $L_\perp$ is the size of the transverse direction with $L_\perp\ll \frac{\hbar}{mc}$ so that excitations with transverse momenta are frozen, eq. (\ref{duedue}) can be rewritten as
\be \label{duetre}
\left(\Box^{(2)} - V \right)\hat \theta^{(2)}=0\ ,
\ee
where $V$ is given by
\be
V= -\frac{1}{2} \left(1-\frac{v^2}{c^2}\right) \frac{d^2c}{dx^2} +\left(\frac{1}{4c}-\frac{5v^2}{4c^3} \right)\left(\frac{dc}{dx}\right)^2  \;   \label{duequattro}
\ee
and $\Box^{(2)}$ is the two dimensional (2D) d'Alembertian associated with the 2D section of the line element (\ref{dueuno}), namely
\be \label{duecinque}
ds^2=\frac{n}{m}\left[-\frac{c(x)^2-v^2}{c(x)}dt^2+\frac{c(x)}{c(x)^2-v^2}dx^2\right]\; .
\ee
Here we have introduced a ``Schwarzschild" time $t$ such that
\be\label{duesei}
    t=T-\int^x dy \frac{v}{c(y)^2-v^2} \;.\ee
 By considering the coordinate $x^*$ , given by
 \be \label{dueotto}
 x^*=\int^x \frac{c(y)dy}{c(y)^2-v^2}\ ,
 \ee
 we can rewrite (\ref{duetre}) in the form
 \be \label{eqep}
\left(\frac{\partial^2}{\partial t^2}+\frac{\partial}{\partial x^{*2}}+V_{\rm eff}\right)\theta^{(2)}=0\ ,
\ee
with effective potential
\be \label{ep}
V_{\rm eff}=\frac{c^2-v^2}{c}V. 
\ee
  
\begin{figure}[h]
\centering
\includegraphics[width=4.5in] {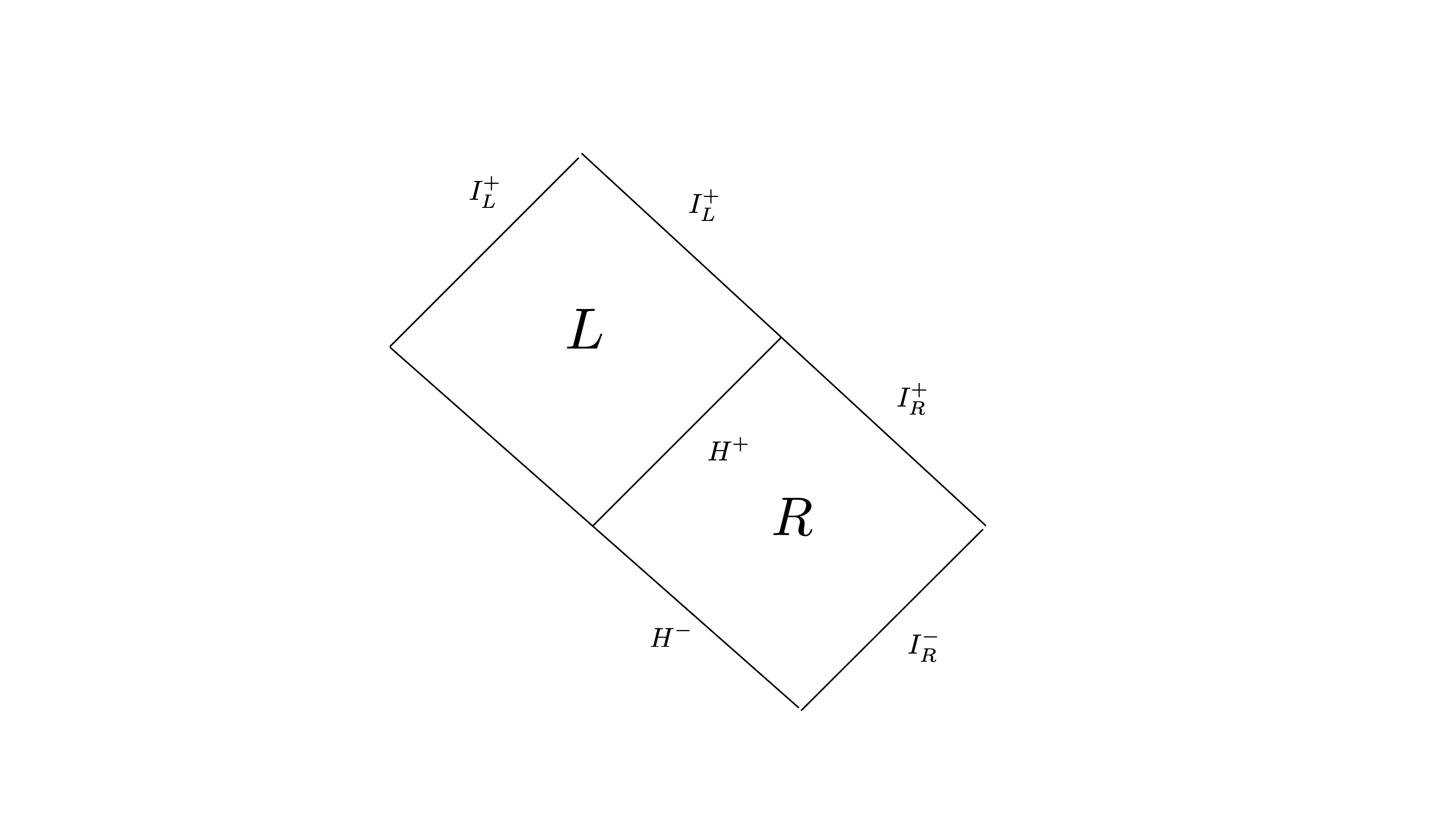}
\caption{Penrose diagram of the spacetime of eq. (\ref{duecinque}). $H^+$ is the future horizon, $H^-$ the past one, $I^\pm_{R,L}$ are null infinities. $R$ is the region outside the horizon, $L$ is the one inside.}
\label{figuno}
\end{figure}

The BEC flows along the $\hat x$ direction from right to left (i.e. $\vec v = -v_0\hat x$, with $v_0>0$). By varying $c(x)$ one can engineer the flow so that it is subsonic for $x>0$ ($R$ region) and supersonic for $x<0$ ($L$ region). $x=0$ is the sonic horizon. $c(x)$ can be further assumed to approach constant values asymptotically; namely $\lim_{x\to +\infty} c(x)=c_R>v_0$ and $\lim_{x\to -\infty} c(x)=c_L<v_0$. Note that the effective potential $V_{\rm eff}$ (\ref{ep}) vanishes asymptotically and on the horizon.
The Penrose diagram of the spacetime is depicted in Fig. (\ref{figuno}).

For further use we introduce the retarded ($u$) and advanced ($v$) null Eddington-Finkelstein like coordinates as
\bea u &=& t-x^* \label{duesettea} \ , \\
v &=& t+x^*\ , \label{duesetteb} \eea
and the Kruskal one
\be
 U = \pm \frac{1}{\kappa} e^{-\kappa u} \ , \label{duenovea} \ee
 where $\kappa$ is the surface gravity of the horizon
 \be \kappa=   \left. \frac{dc}{dx}\right|_{x=0} \label{duedieci} \ee
 and in eq. (\ref{duenovea}) the plus sign is for the $L$ region and the minus sign is for the $R$ region.
 
\begin{figure}[h]
\centering
\includegraphics[width=4.5in] {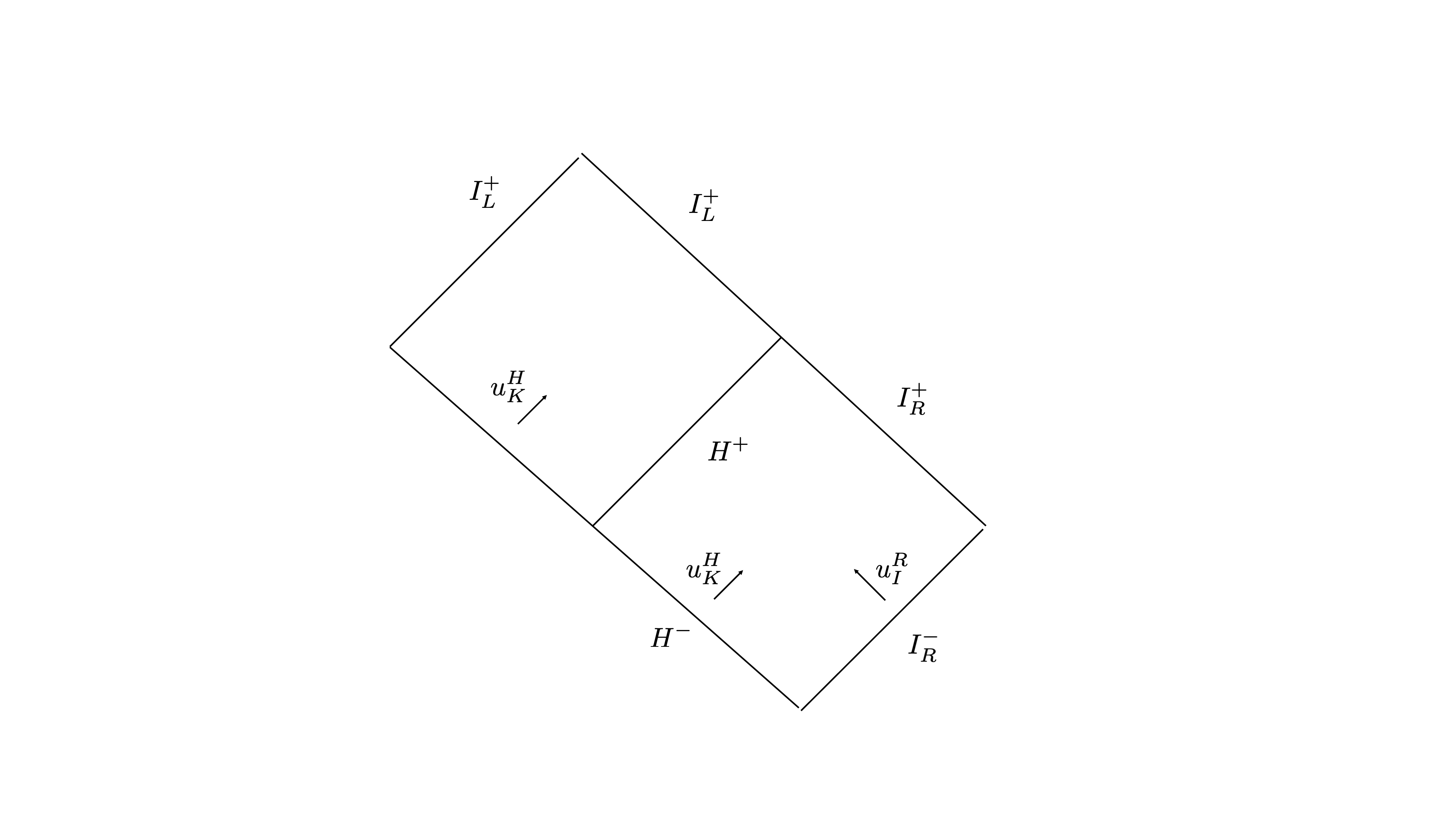}
\caption{Modes for the Unruh state.}
\label{figdue}
\end{figure}

 The quantum state of our field $\hat\theta^{(2)}$, as is well known, can be approximated at late times after the formation of the BH by the Unruh state $|U\rangle$ \cite{unruh76}.
 This corresponds to an expansion of the quantum operator as
 \be \label{dueundici}
\hat{\theta}^{(2)}=\int_0^\infty d\omega_K \left[\hat a_K(\omega_K) u_H^{K}+h.c.
\right]+\int_0^\infty d\omega \left[\hat a_I(\omega) u_I^{R}+h.c.
\right]\ ,
\ee
 where the form of the mode $u_I^R$ on $I_R^-$ is
  \be \label{duedodici}
 u_I^{R}(\omega, x)=\frac{e^{-i\omega v}}{\sqrt{4\pi \omega}}\ ,
 \ee
 while that for the mode $u_H^K$ on $H^-$ is
  \be \label{duetredici}
 u_H^K(\omega_K, x)=\frac{e^{-i\omega_K U}}{\sqrt{4\pi \omega_K}}\ .
 \ee
 The Unruh state $|U\rangle$ is defined as
 \be \hat a_K(\omega_K)|U\rangle = 0 = \hat a_I(w) |U\rangle \label{duequattordici}\   \ee
 for every $\omega, \omega_K$. In Fig. (\ref{figdue}) we illustrate these modes on the Cauchy surface
 $H^-\cup I_R^-$.
 
 Note that while $u_I^R$ is a positive frequency mode with respect to Schwarzschild time, $u_H^K$
 is positive frequency with respect to Kruskal time.
 The Unruh state describes a situation in which one has no incoming radiation on $H^- \cup I_R^-$ , while at $I_R^+$ one has a thermal flux at the Hawking temperature $\displaystyle{T_H=\frac{\hbar\kappa}{2\pi k_B}}$, where $k_B$ is Boltzmann's constant.
 
\begin{figure}[h]
\centering
\includegraphics[width=4.5in] {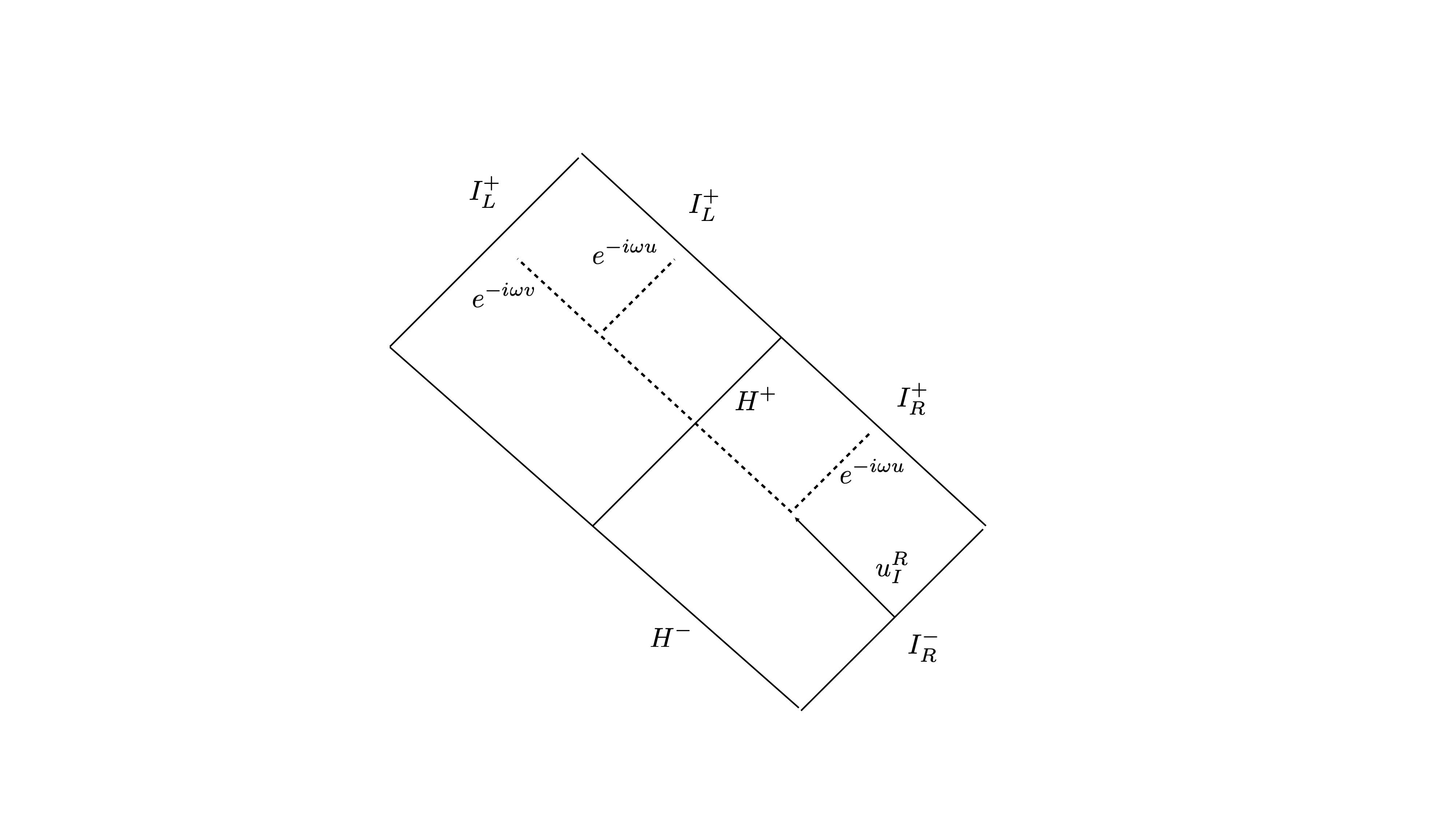}
\caption{Backscattering of the $u_I^R$ modes.}
\label{figtre}
\end{figure}

\begin{figure}[h]
\centering
\includegraphics[width=4.5in] {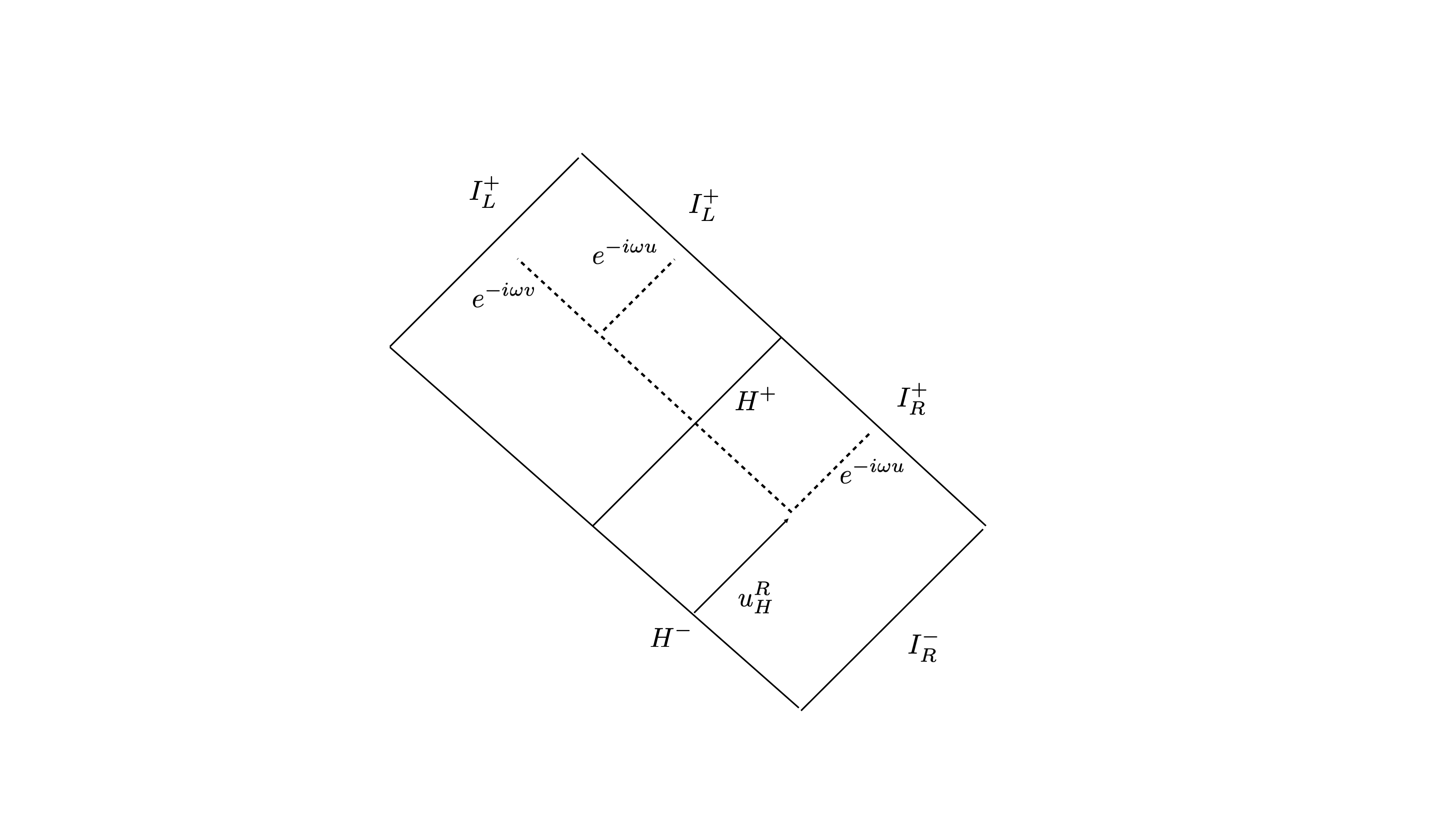}
\caption{Backscattering of the $u_H^R$ modes.}
\label{figquattro}
\end{figure}

\begin{figure}[h]
\centering
\includegraphics[width=4.5in] {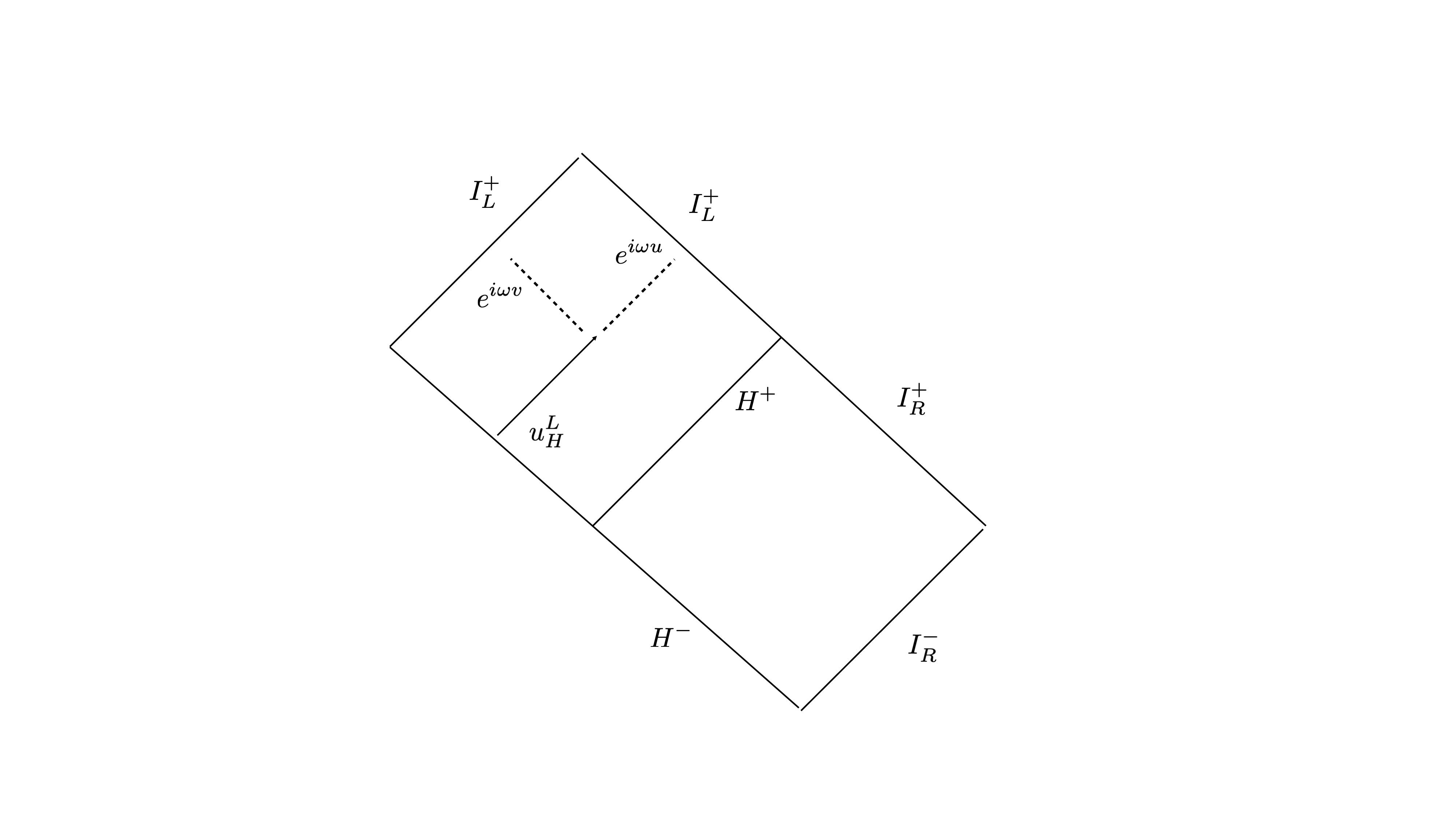}
\caption{Backscattering of the $u_H^L$ modes.}
\label{figcinque}
\end{figure}

 It is convenient for the calculations of the 2-point function to express the modes $u_H^K$ in terms of modes (Boulware modes \cite{bo}) that on $H^-$ behave as
 \bea
  u_H^{R} &=& \frac{e^{-i\omega u}}{\sqrt{4\pi \omega}}\ , x>0, \label{duequindicia} \\
  u_H^{L} &=& \frac{e^{i\omega u}}{\sqrt{4\pi \omega}}\ , x<0 \label{duequindicib} \ . \eea
  Note the $+$ sign in front of the exponential in eq. (\ref{duequindicib}). $u_H^L$ has negative frequency (while having positive norm), it is associated with the negative (Killing) energy partners.
  We have
  \be \label{duesedici}
 u_H^{K}(\omega_K,x)=\int_0^\infty d\omega \left[\alpha_{\omega_K \omega}^R u_H^{R}+\beta^{R}_{\omega_K \omega} u_H^{R*}\right]+ R \leftrightarrow L \ .
 \ee
  The Bogoliubov coefficients are given in Ref. \cite{paper2013} \footnote{In Ref. \cite{paper2013} there is a misprint in eqs. (4.14b)-(4.14e). $\omega_K$ should be replaced by $\frac{\omega_K}{\kappa}$.} and are summarised in Appendix \ref{appendixA}.
  Because of the presence of the effective potential $V_{\rm eff}$ in (\ref{ep}), 
  the incoming modes will be modified from their asymptotic forms in Eqs. (\ref{duedodici}), (\ref{duequindicia}), (\ref{duequindicib}) 
  due to backscattering effects. In Ref. \cite{paper1}, $V$ was neglected and the modes maintained their expressions (\ref{duedodici}), (\ref{duequindicia}), (\ref{duequindicib}) throughout the space-time. In Figs. (\ref{figtre}) - (\ref{figcinque}) we schematically describe the backscattering of each mode.
 
In any case, since $V_{\rm eff}$ vanishes asymptotically at future null infinity each incoming mode will be a linear combination of $e^{-i\omega v}$ and $e^{\pm i\omega u}$.

We will consider two profiles of the sound velocity. The first one has been proposed in a numerical simulation fully based on the Bogoliubov theory of a BEC \cite{cfrbf}  which confirmed the presence of a peak in the in-out correlation function as predicted by \cite{paper1} using only QFT in curved space methods. It is
\bea  c(x)&=&\sqrt{c_L^2 + \frac{1}{2}(c_R^2-c_L^2)\left[1+\frac{2}{\pi}\tan^{-1}\left(\frac{x+b}{\sigma_v}\right)\right]}  \;, \label{duediciassette} \\
      b &=& \sigma_v \tan \left[\frac{\pi}{c_R^2-c_L^2} \left( v_0^2 - \frac{1}{2} (c_R^2+c_L^2)\right) \right] \;, \label{duediciotto} \eea
where $\sigma_v$ is an arbitrary positive constant. The horizon is at $x=0$. The surface gravity for this profile is
 \be \kappa = \left. \frac{d c}{dx} \right \vert_{x=0} = \frac{1}{2 \pi v_0 \sigma_v}(c_R^2-c_L^2) \sin^2 \left[\frac{\pi(c_R^2-v_0^2)}{c_R^2-c_L^2}\right] \label{kappa-2013-profile}\ .  \ee
 We call this the ``original" profile.
In this case the equation for the modes eq. (\ref{duetre}) has to be solved numerically. The explicit construction of the modes has been given in Ref. \cite{paper2013}, written in collaboration with Renaud Parentani, where all the details can be found.
The second profile, which we call ``analytical", is
\be c(x) =  \frac{c_R}{\sqrt{1 + \left(\frac{c_R^2}{v_0^2} - 1 \right) \exp \left[-\frac{2 \frac{c_R^2}{v_0^2} k x}{(\frac{c_R^2}{v_0^2} - 1)}\right]}} \theta(x) +
 \frac{c_L}{\sqrt{1 - \left(1 -\frac{c_L^2}{v_0^2} \right) \exp \left[-\frac{2 k \frac{c_L^2}{v_0^2} x}{\left(\frac{c_L^2}{v_0^2}-1\right)}\right]}}\theta(-x)  \;, \label{duediciannove}
\ee    
where $x=0$ is the horizon and $k$ is a positive constant of dimension $L^{-1}$ and the corresponding surface gravity is
\be
\kappa=  \left. \frac{dc}{dx} \right|_{x=0}=k \, v_0\ . \label{kappa-analytical-profile} \ee
This profile has been introduced in Ref. \cite{fba2016}. The advantage of this profile is that the modes can be computed analytically in terms of hypergeometric functions.

 \section{Correlation functions}
  
  As mentioned in the Introduction, the only experimental support for the existence of Hawking-like radiation in an acoustic BH formed by a BEC comes from the observation of a correlation band appearing in the in-out (one point inside the horizon and the other outside) equal time density-density correlation function, in 
  agreement with the theoretical prediction \cite{paper1}.

  
 Defining the operator $\hat n$ as the quantum density fluctuation on top of the condensate, the density-density correlation function is
 \be
G_2\left(T, x;T^\prime, x^\prime\right)=
\langle U|\hat{n}(T, x)\hat{n}(T^\prime, x^\prime)|U \rangle\ . \label{treuno}
\ee
 In the hydrodynamical approximation we have~\cite{Barcelo:2005fc}  
 \be
\hat{n}= \frac{\hbar n}{mc^2}\left[v_0\partial_x\hat{\theta}-\partial_T  \hat{\theta}\right] \;. \label{eqtredue}
\ee
One finds that
\be
G_2\left(T, x;T, x^\prime \right) =
\frac{\hbar n}{2 m L_\perp^2 c^{2}(x)c^{2}(x^\prime)} D \sqrt{c(x)c(x')}\langle U| \left\{ \hat{\theta}^{(2)}(t, x), \hat{\theta}^{(2)}(t^\prime, x^\prime)\right\}|U\rangle \ , \label{tretre} \ee
where
\be
D \equiv \partial_T\partial_{T^\prime}-v_0\partial_x\partial_{T^\prime}-v_0\partial_T\partial_{x^\prime}+v_0^2\partial_x\partial_{x^\prime}
 \ . \label{trequattro} \ee
 Using eqs. (\ref{duedodici}) and (\ref{duesedici}) and integrating over $\omega_K$  (see \cite{paper2013}) the two-point function entering (\ref{tretre}) can be written as~\cite{paper1}
 \be \langle U| \left\{ \hat{\theta}^{(2)}(t, x), \hat{\theta}^{(2)}(t^\prime, x^\prime) \right\} |U\rangle = I + J\ ,  \label{trecinque}\ee
 where
\bea
 I &=& \int_0^\infty d \omega \frac{1}{\sinh\left(\frac{\pi \omega}{\kappa}\right)} \left\{ u^{L}_H(\omega,t,x) \, u^{R}_H(\omega,t',x') +
   u^{L *}_H(\omega,t,x) \, u^{R *}_H(\omega,t',x') \right. \nonumber \\
   & & \;\;\; \left. + u^{R}_H(\omega,t,x) \, u^{L}_H(\omega,t',x') +
   u^{R *}_H(\omega,t,x) \, u^{L *}_H(\omega,t',x') \right. \nonumber \\
   & & \;\; \left.  + \cosh \left(\frac{\pi \omega}{\kappa}\right) \left[ u^{L}_H(\omega,t,x) \, u^{L *}_H(\omega,t',x')
   + u^{L *}_H(\omega,t,x) \, u^{L}_H(\omega,t',x') \right. \right. \nonumber \\
  & & \;\;\; \left. \left. + u^{R}_H(\omega,t,x) \, u^{R *}_H(\omega,t',x') + u^{R *}_H(\omega,t,x) \, u^{R }_H(\omega,t',x') \right]
   \right\}     \;, \label{tresei} \\
   J &=& \int_0^\infty d \omega \, \left[ u^{R}_I(\omega,t,x) \, u^{{R} \, *}_I(\omega,t',x')
   +  u^{{R} \, *}_I(\omega,t,x) \, u^{R}_I(\omega,t',x') \right] \; \label{tresette}
    \eea    
 and the relation between Schwarzschild time $t$ and $T$ is given by  eq. (\ref{duesei}).
 
\begin{figure}[h]
\centering
\includegraphics[width=3in] {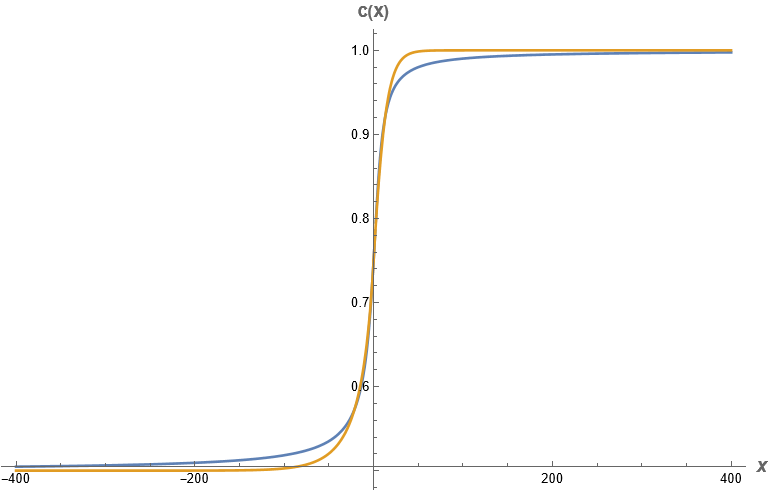}
\caption{Plot for the profiles of Eqs. (\ref{duediciassette}) (blue curve), (\ref{duediciannove}) (orange curve) arranging them to have the same surface gravity and asymptotic behaviour.}
\label{figsei}
\end{figure}

\begin{figure}[h!]
\centering
\includegraphics[width=3in] {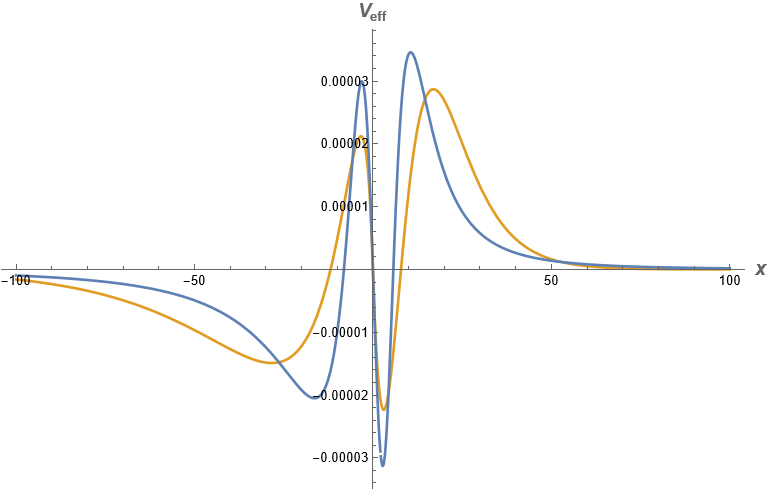}
\caption{Plot of the effective potential (\ref{ep}), (\ref{duequattro}) for the profiles of eqs. (\ref{duediciassette}) (blue curve), (\ref{duediciannove}) (orange curve) with the same asymptotic behaviour and surface gravity..}
\label{figsette}
\end{figure}

\begin{figure}[h!]
\includegraphics[scale=.32]{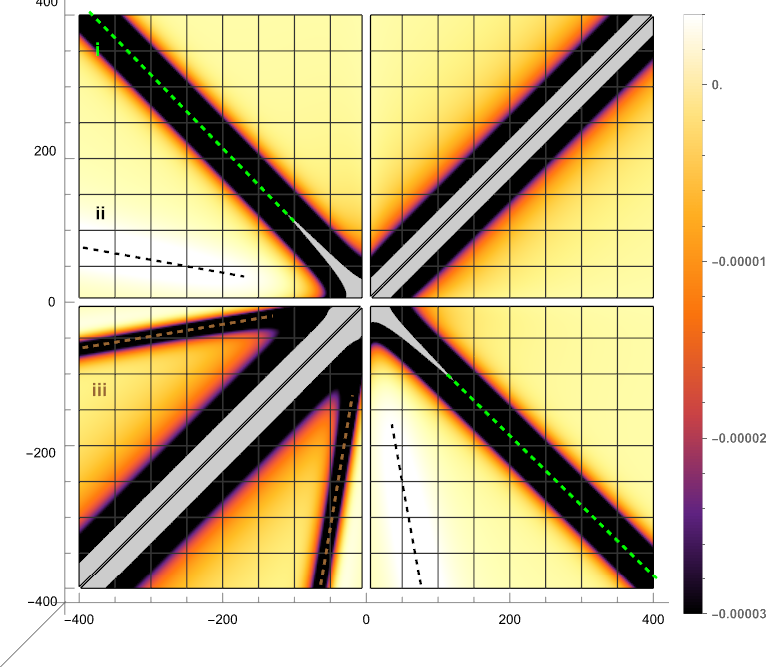}
\includegraphics[scale=.32]{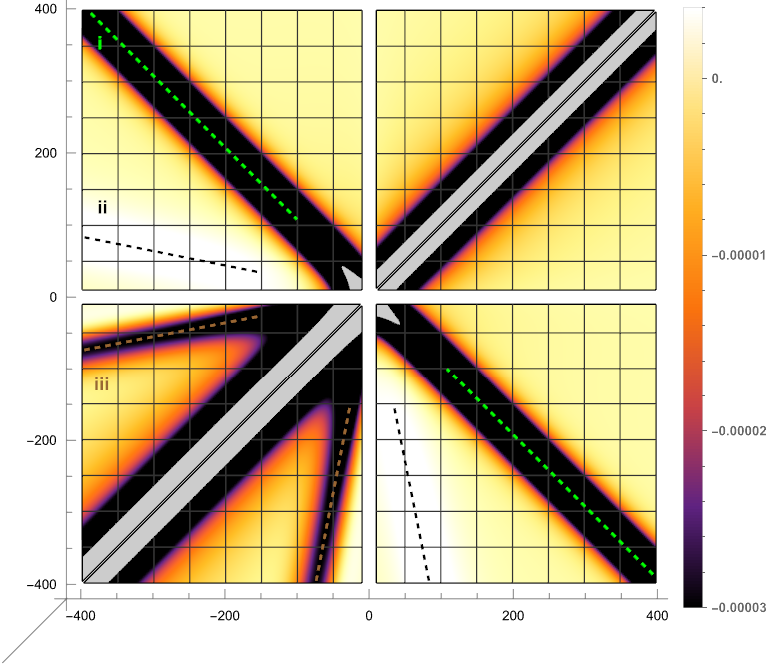}
\caption{\label{Fig:DDCF Full} Density-density correlation function for the two sound speed profiles. Left: Original profile (\ref{duediciassette}). Right:  Result using the analytical profile (\ref{duediciannove}). Each figure is oriented so that the bottom left quadrant where both points are in the interior and the upper right quadrant where both points are in the exterior of the analog black hole. 
 The green-dashed lines highlight the main band (i) by showing the locations of two parts of its negative correlation peak.  The black-dashed and brown-dashed lines do the same for the secondary positive band (ii) and the secondary negative band (iii) respectively.  Regions shaded in gray are outside of the ranges of the plots.}
\label{figotto}
\end{figure}

\begin{figure}[h]
\centering
\includegraphics[width=3in] {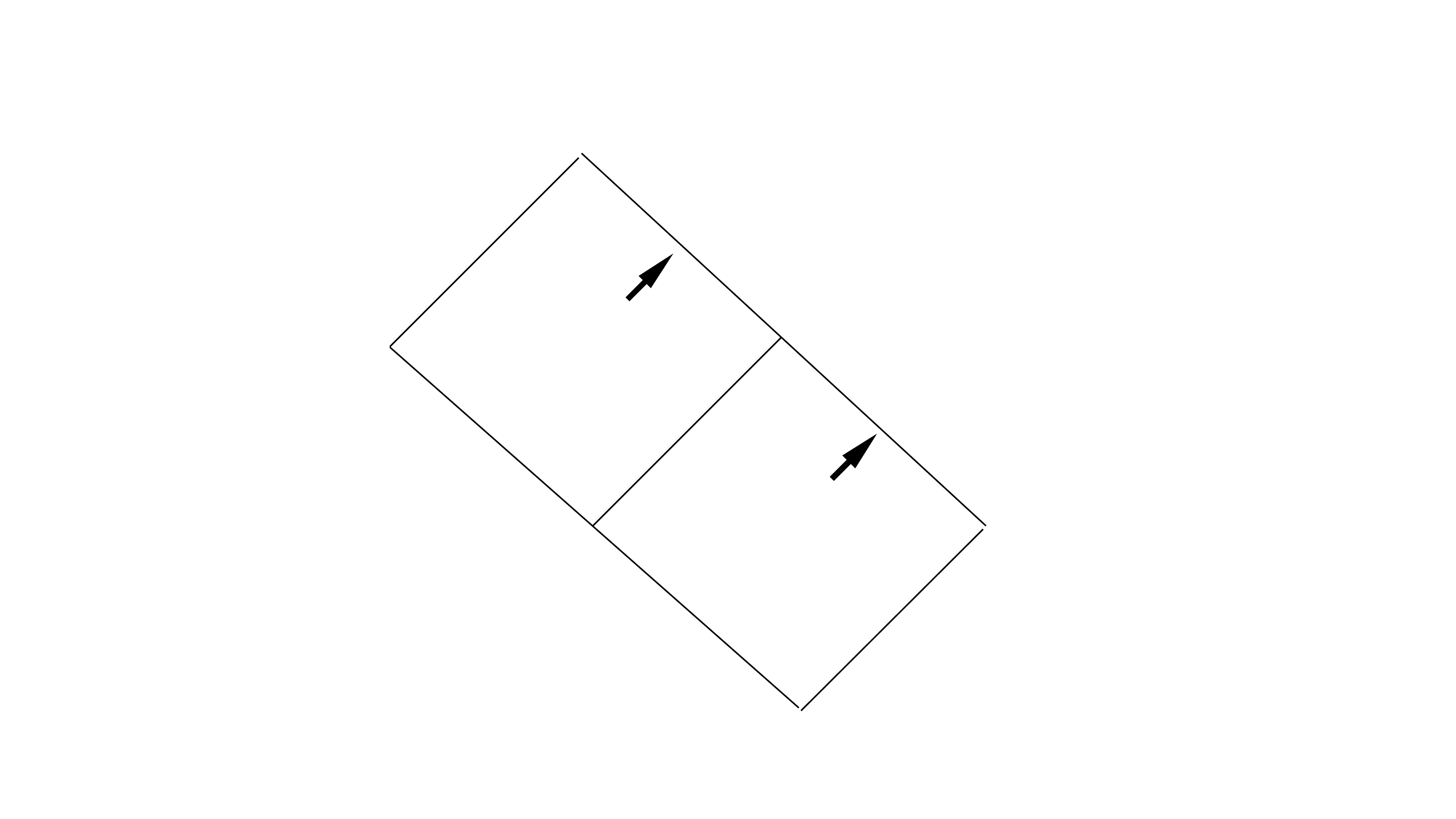}
\caption{Correlations between $u,u'$ modes on both sides of the horizon.}
\label{Figi}
\end{figure}

\begin{figure}[h]
\centering
\includegraphics[width=3in] {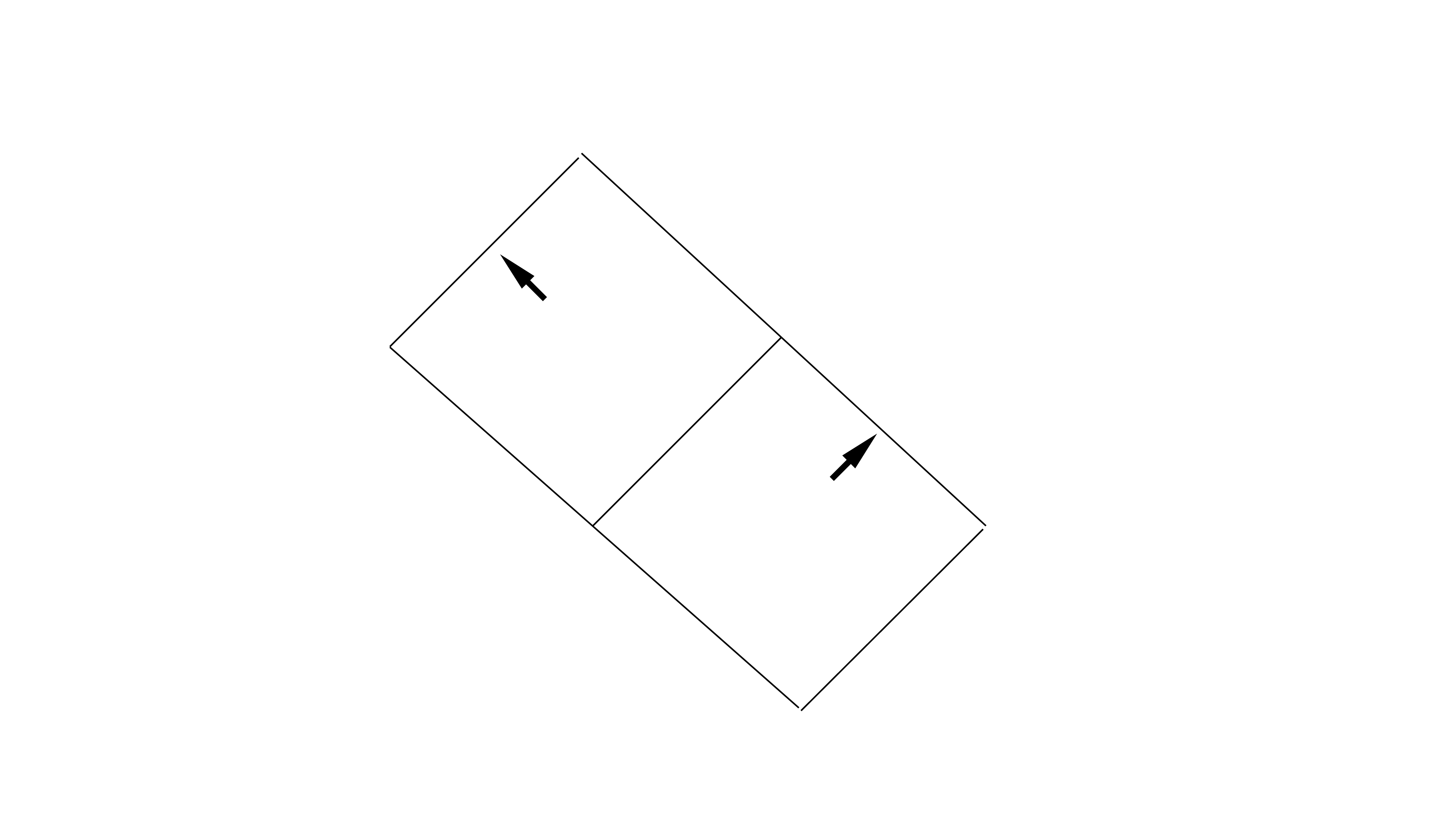}
\caption{Correlation between $u,v'$ modes  modes across the horizon.}
\label{Figii}
\end{figure}

\begin{figure}[h]
\centering
\includegraphics[width=3in] {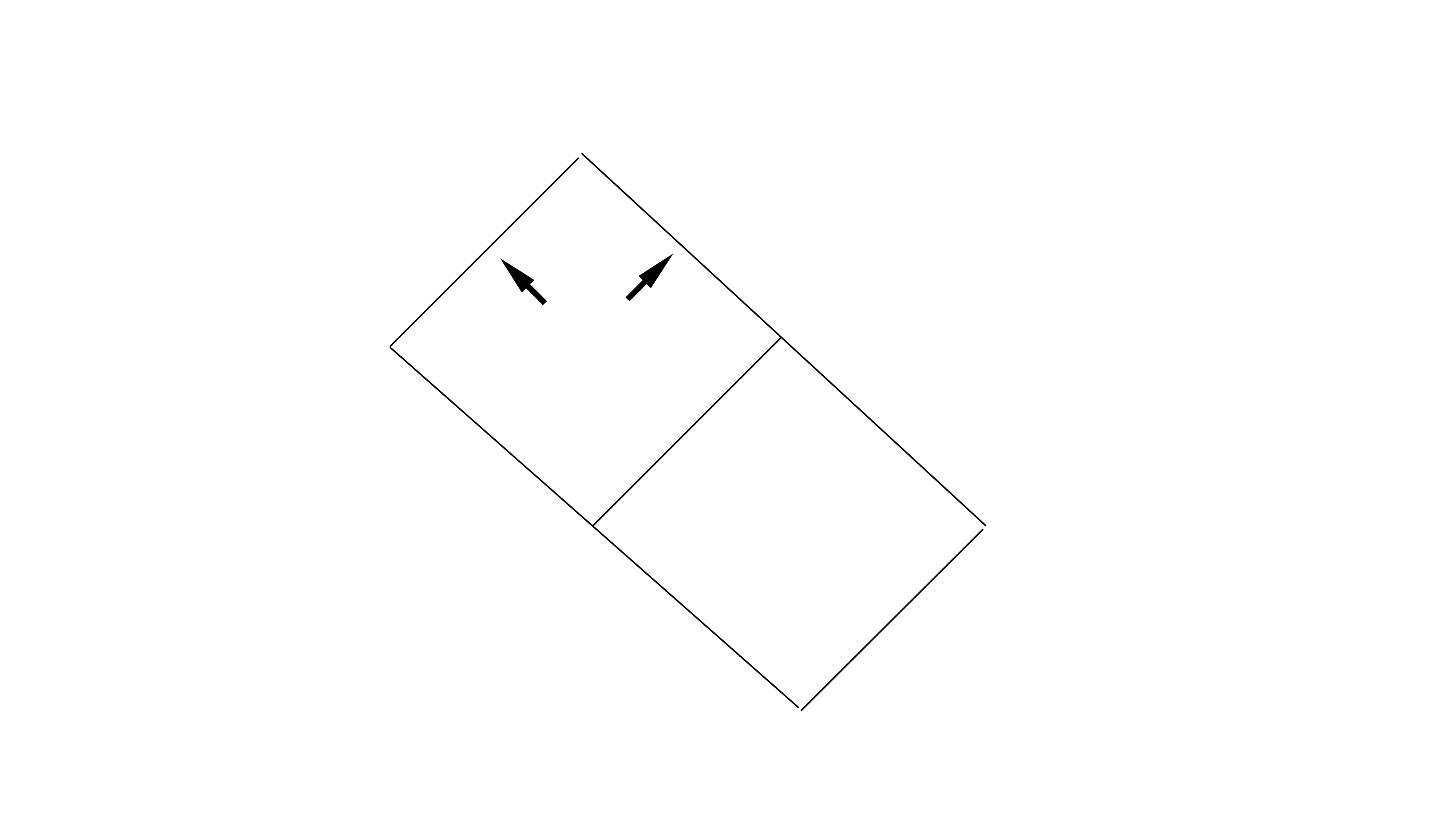}
\caption{Correlation between $u,v'$ modes  modes inside the horizon.}
\label{Figiii}
\end{figure}

 If one neglects the effective potential in eq. (\ref{eqep}), the modes maintain the form given by eqs.
  (\ref{duedodici}), (\ref{duequindicia}), (\ref{duequindicib}) throughout the entire space-time. In this case one can obtain an analytical expression for $G_2(T,x;T',x')$ which, taking the point $x$ in the asymptotic $R$ region where $c(x)\sim c_R$ and the point $x'$ in the asymptotic $L$ region where $c(x)\sim c_L$, can be written as
\bea G_2(T,x;T,x') &=&  \frac{\hbar n}{2 m L_\perp^2 c_R^{3/2}c_L^{3/2}} \Big\{
  -\frac{1}{(c_R-v_0)(v_0-c_L)}\frac{\kappa^2}{\cosh^2 \frac{\kappa}{2}(u-u')}\nonumber \\ &+& \frac{1}{(c_R+v_0)(c_L+v_0)}\frac{1}{(v-v')^2}\Big\}  \ . \label{treotto} \eea
  We see that this function has a negative minimum peaked along $u=u'$ which corresponds (in the geometrical optics approximation) to the trajectory of the Hawking quanta ($u=const$) and its partner ($u'=const$). 
Beside this no other structure is present. This feature is the one observed by Steinhauer's group \cite{jeff2019, jeff2021}.
We see that the no-backscattering asymptotic correlation function (\ref{treotto}) has the same form for all profiles having the same $c_R,c_L,v_0$ and surface gravity $\kappa$; in particular the height, width, and location of the minima are identical. We shall impose this to be the case for our two profiles introduced in the previous section.  In Fig. (\ref{figsei}) we have plotted the two profiles  for $v_0=\frac{3}{4}, c_L=\frac{1}{2}, c_R=1, \sigma_v=8$ and numerically matched the two surface gravities.  
Even with all these parameters matched, there are noticeable differences in the comparison of the sound profiles. This leads to differences in the scattering of the modes as can be seen in Fig. (\ref{figsette}), where the effective potential for the two profiles is plotted.  The extrema of the potential appear to be higher and narrower for the original profile.  All this has a significant signature in the correlation functions as we shall see.
In Fig. (\ref{figotto}) we have represented the correlation function Eq. (\ref{tretre}) at equal time $T=T'$ for the original and the analytical profile respectively.
This function is symmetric under the exchange $x\leftrightarrow x'$ and diverges 
when the points come together and thus the region $x=x'$ is cutoff for this reason. In each figure one can clearly see the large negative correlation band, i), when one point is in the interior and the other point is in the exterior region. This is the one predicted by the no-backscattering asymptotic expression (\ref{treotto})  and corresponds to the correlation between the modes depicted in Fig. (\ref{Figi}). One can also see two much smaller bands predicted by R. Parentani: a positive one, labeled (ii), when one point is inside the horizon and the other outside (correlations between the modes represented in Fig. (\ref{Figii})) and a negative one when both points are inside the horizon, labeled (iii) (correlation corresponding to the modes in Fig. (\ref{Figiii})).
These two secondary correlation bands exist because of the backscattering of the modes.

\begin{figure}[!ht]
\includegraphics[scale=.5]{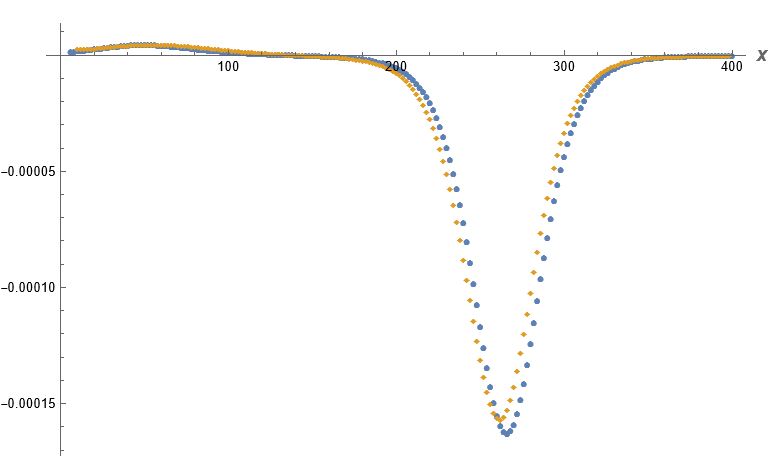}
\caption{\label{Fig:FullInOutPlots1} The density-density correlation function at $x^\prime=-250$ when one point is in the interior and one point in in the exterior of the analog black hole. Blue dot: Original profile (\ref{duediciassette}). Orange diamond:  Analytic profile (\ref{duediciannove}). }
\label{fignove}
\end{figure}

\begin{figure}[!ht]
\includegraphics[scale=.5]{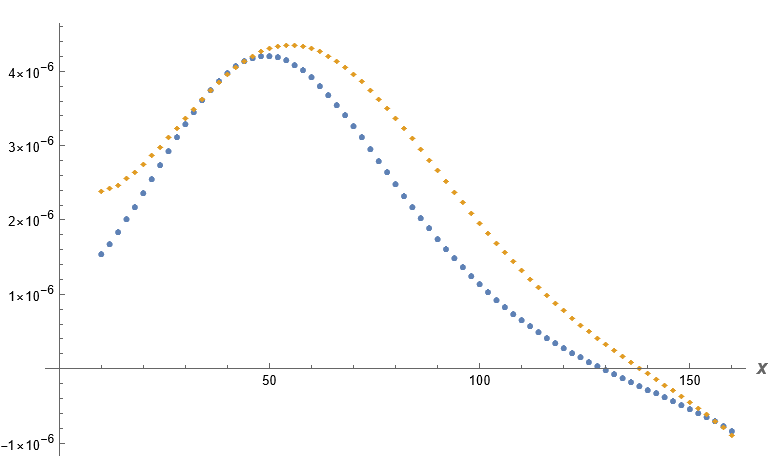}
\caption{\label{Fig:SmallPeakInOut} A comparison of the secondary peak at $x^\prime=-250$ when one point is in the interior and one point in in the exterior of the analog black hole.  Blue dot: Original profile (\ref{duediciassette}). Orange diamond:  Analytic profile (\ref{duediciannove}).}
\label{figdieci}
\end{figure}

To see the differences between the two profiles, we have taken in Fig. (\ref{fignove}) a slice at $x'=-250$ of the in-out region ($x'<0,x>0$) of Fig. (\ref{figotto}). In this and in the following figures the extrema of the correlations bands will appear as peaks.
The large main peak corresponding to the i) correlation band is clearly visible but the peak height and location are offset for the two profiles. The main negative peak for the analytic profile appears slightly smaller and shifted to the left as compared to that of the original one. The opposite occurs for the smaller secondary peak corresponding to the band ii) as seen in Fig. (\ref{figdieci}),
where we have magnified the scale to better appreciate this point.
\begin{figure}[!ht]
\includegraphics[scale=.5]{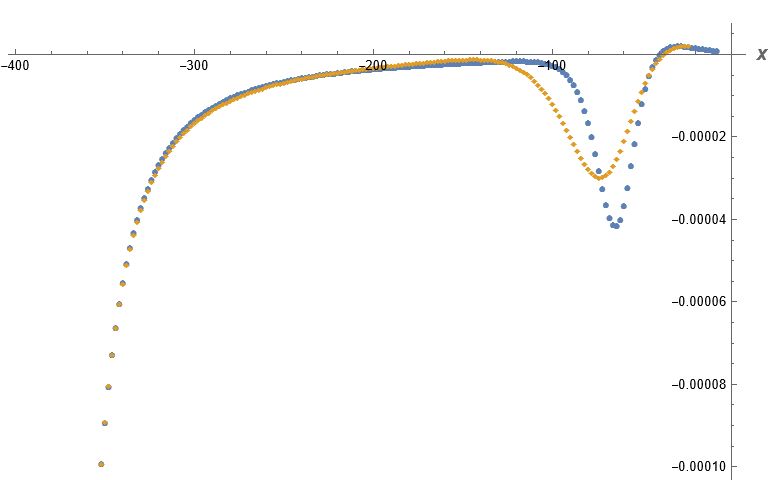}
\caption{\label{Fig:FullPlotInIn2} Density-density correlation function for both points in the interior at a fixed ${x^\prime =400}$. Blue dot: Original profile (\ref{duediciassette}). Orange diamond:  Analytic profile (\ref{duediciannove}).   }
\label{figundici}
\end{figure}
More striking is the relative difference appearing in the negative peak corresponding to the band iii) in the in-in region, see Fig. (\ref{figundici}).
\begin{figure}[!h]
\includegraphics[scale=.33]{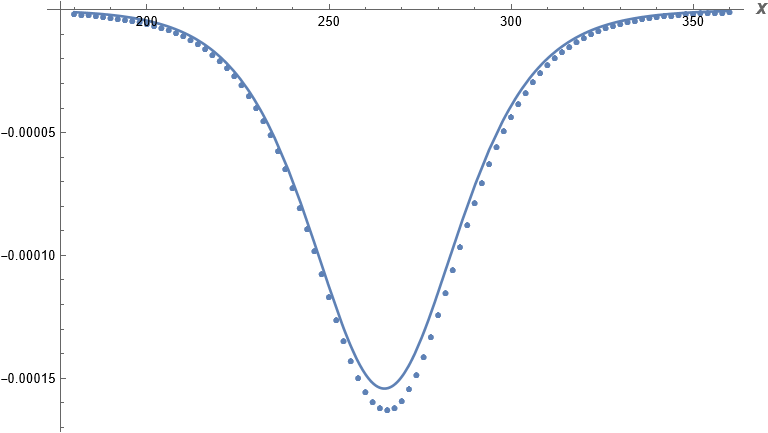}
\includegraphics[scale=.33]{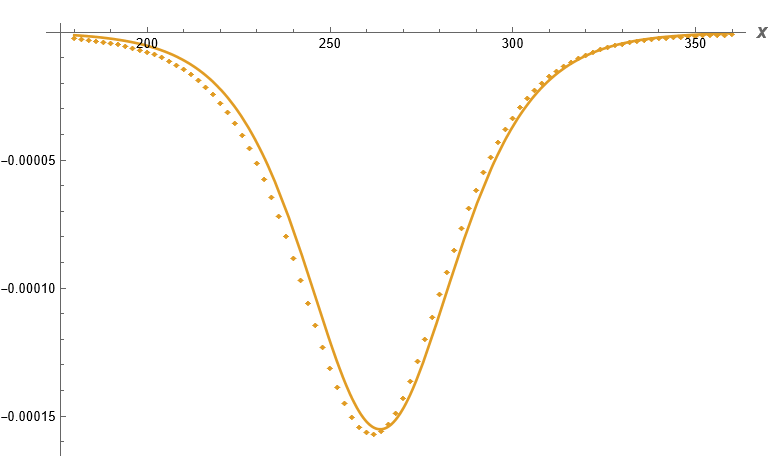}
\caption{\label{Fig:MainPeakInOut}A comparison of the main peak at $x^\prime=-250$ when one point is in the interior and one point in in the exterior of the analog black hole. Left: Blue(Solid): Curve for the analytic expression~\eqref{treotto}. Blue dot: Numeric data for original profile (\ref{duediciassette}). Right: Orange (solid): Curve for the analytic expression~\eqref{treotto}. Orange diamond:  Numeric result for analytic profile (\ref{duediciannove}).}
\label{figdodici}
\end{figure}
The backscattering also affects the main peak. In Fig. (\ref{figdodici}) a comparison is made for the two profiles with the no-backscattering approximation eq. (\ref{treotto}). The differences are more significant for the original profile.

Finally, it is interesting to compare the ratios of the heights of each of the two minor peaks to that of the main peak.   The ratio of the height of the positive minor peak with one point inside and one point outside the horizon to that of the main peak for the original profile is $0.0293$.   For the analytical profile it is $0.0302$.  The ratio of the height of the negative peak when both points are inside the horizon to the main peak for the original profile is $0.261$.  For the analytical profile it is $0.191$.  In both cases there is agreement in the first digit only, so differences in the profiles lead to relatively significant differences between the two profiles.

\section{Scaling}

There is a scaling related to the surface gravity $\kappa$ that works for both sound speed profiles used in this paper, (\ref{duediciassette}) and (\ref{duediciannove}).  It is
\bes \bea \bar{\w} &=& \frac{\w}{\kappa} \;, \\
    \tau &=& \kappa t  \;,  \\
     \xi &=& \kappa x  \;. \label{xi-def} \\
\eea \label{scaled-quantities} \ees
It is easy to see that for this scaling both sound speed profiles, written in terms of $\xi$, are independent of $\kappa$.

In general, for any sound speed profile that, when written in terms of $\xi$, is independent of $\kappa$ one can substitute ~\eqref{scaled-quantities} into the Boulware and Kruskal modes of (\ref{dueundici}) and show that they both scale as $\kappa^{-1/2}$. Using these results, one can show that the two-point function (\ref{trecinque}) is independent of $\kappa$. Then one finally has from (\ref{tretre}) that

\be G_2(T,x;T',x') = \kappa^2 \; G_2(\bar{T},\xi;\bar{T}',\xi') \label{scaled-G2}  \;. \ee

\begin{figure}[!h]
\includegraphics[scale=.32]{Fig8Left.png}
\includegraphics[scale=.32]{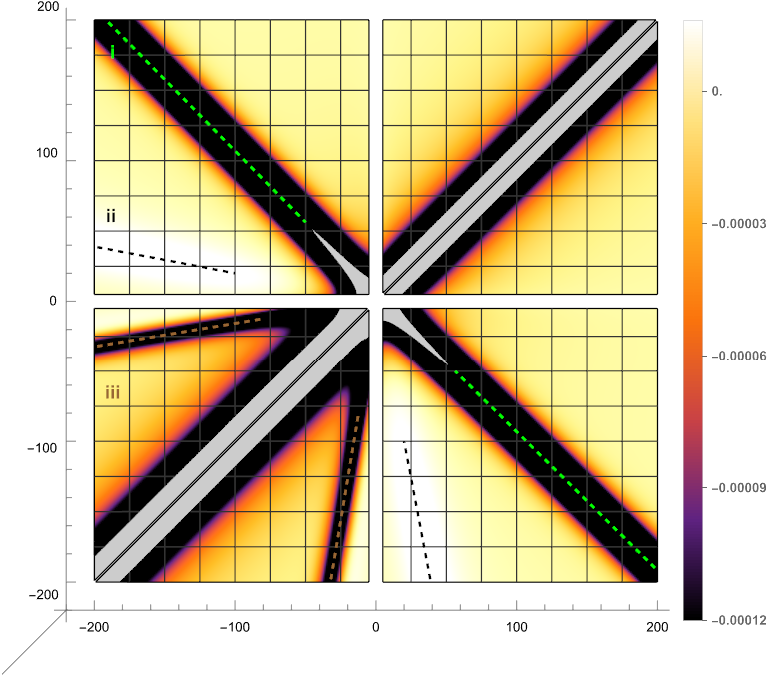}
\caption{\label{Fig:ScalingComparisonFullPlots} Scaling present in the density-density correlation function: Left: Result using the original profile (\ref{duediciassette}) with $\sigma_v=8$ with axis limits at $-400\leq x \leq 400 $. Right:  Result using the Original profile \ref{duediciassette} with $\sigma_v=4$    with axis limits at $-200\leq x \leq 200 $. 
The green-dashed lines highlight the main band (i) by showing the locations of two parts of its negative correlation peak.  The black-dashed and brown-dashed lines do the same for the secondary positive band (ii) and the secondary negative band (iii) respectively.  Regions shaded in gray are outside of the ranges of the plots.} 
\label{figtredici}
\end{figure}

\begin{figure}[!h]
\includegraphics[scale=.35]{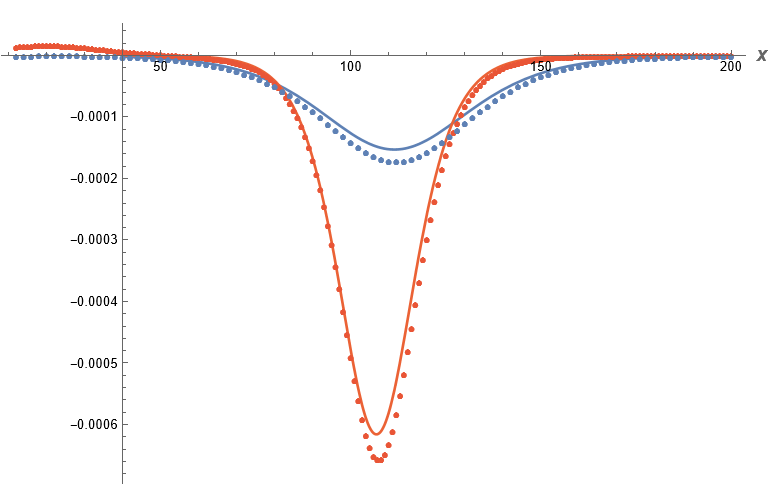}
\caption{\label{Fig:ScalingComparisonMainPeak}  
Comparison of the main peak in analytic and numerical results for  with $\sigma_v=8$ and $\sigma_v=4$ along a slice at $x=-100$. Blue  (solid ): Theoretical curve (\ref{treotto}) for the original profile (\ref{duediciassette}) with $\sigma_v=8$.  Blue diamond: Numerical result for the original profile (\ref{duediciassette}) with $\sigma_v=8$. Red (solid): Theoretical curve for original profile (\ref{duediciassette}) with $\sigma_v=4$. Red square: Numerical result for the original profile (\ref{duediciassette}) with $\sigma_v=4$.     }
\label{figquattordici}
\end{figure}

This means that the heights and depths of the correlation peaks are larger for larger values of $\kappa$.  Since $x = \frac{\xi}{\kappa}$, the widths of the correlation peaks in terms of the space coordinate $x$ are narrower for larger values of $\kappa$. See Figs. (\ref{figtredici}, \ref{figquattordici}).

\section{Conclusions}

While awaiting the direct detection of the thermal phonons, the correlations bands and their related peaks in the density-density correlation function are the major tool to experimentally investigate the analogue of Hawking radiation in a sonic BH formed by a BEC. Of the three characteristic bands only one, the main one predicted in \cite{paper1}, has been observed so far.  The detection of the other much smaller two, whose existence was predicted by R. Parentani, represents the next challenge for the experimentalists in this field.

In this paper we have investigated the details of the three characteristic bands for two specific sound velocity profiles.   The profiles we use have flow speeds that are constant and sound speeds that vary with position.  To date, the experimental profiles have had both flow and sound speeds that vary with position.  However, the three bands are expected to be very robust in the sense that they should occur for any profiles in which the sound and flow speed profiles are effectively one dimensional, result in a single sonic horizon, and approach constant values far from that horizon. 

The goal of this work was to show how sensitive the correlations are to differences in the sound speed profile.  We find that significant differences occur for each of the characteristic bands even though the sound speed profiles have the same asymptotic and horizon limits.  Since the profiles are quite similar, this implies that one would expect a significant amount of sensitivity to the details of the experimental profiles as well. 

Future improvements in the precision of the experiments, including  hopefully  the detection of the secondary  bands, will allow for a more detailed analysis of the experimental results.  This in turn should allow us to determine the extent to which the gravitational analogy and the corresponding quantum field theory in curved space calculations can be trusted to explain the fine structure features of the correlations in the Hawking radiation for a BEC analogue black hole.

\acknowledgments

P. R. A. was supported in part by the National Science Foundation under grant No. PHY-2309186.  Some of the numerical work was done using the WFU DEAC Cluster; we thank
the WFU Provost’s Office and Information Systems Department for their generous support.
A.F. acknowledges partial financial support by the Spanish Grants PID2020-116567GB-C21, PID2023-149560NB-C21 funded by MCIN/AEI/10.13039/501100011033, and by the Severo Ochoa Excellence Grant CEX2023-001292- S.

\begin{appendix}
\section{Bogoliubov coefficients relating the Kruskal and Boulware modes}
\label{appendixA}
We report here the Bogoliubov coefficients appaearing in eq. (\ref{duesedici}),
 \be \label{duesedicibis}
 u_H^{K}(\omega_K,x)=\int_0^\infty d\omega \left[\alpha_{\omega_K \omega}^R u_H^{R}+\beta^{R}_{\omega_K \omega} u_H^{R*}+
 \alpha_{\omega_k \omega}^L u_H^{L}+\beta^{L}_{\omega_k \omega} u_H^{L*}\right] \ ,  
 \ee
\bea
      \alpha^R_{\omega_k \omega} &=& \frac{1}{2 \pi \kappa} \sqrt{\frac{\w}{\w_K}} \Gamma(-i \w/\kappa) \; \left(-i \frac{\w_K}{\kappa} \right)^{i \w/ \kappa}\ , \nonumber \\
\beta^R_{\omega_k \omega} &=& \frac{1}{2 \pi \kappa} \sqrt{\frac{\w}{\w_K}} \Gamma(i \w/\kappa) \; \left(-i \frac{\w_K}{\kappa} \right)^{-i \w/ \kappa}\ , \nonumber \\
      \alpha^L_{\omega_k \omega} &=& \frac{1}{2 \pi \kappa} \sqrt{\frac{\w}{\w_K}} \Gamma(i \w/\kappa) \; \left(i \frac{\w_K}{\kappa} \right)^{-i \w/ \kappa}\ , \nonumber \\
\beta^L_{\omega_k \omega} &=& \frac{1}{2 \pi \kappa} \sqrt{\frac{\w}{\w_K}} \Gamma(-i \w/\kappa) \; \left(i \frac{\w_K}{\kappa} \right)^{i \w/ \kappa} \;.   \label{alphas-betas}
      \eea
\end{appendix}

  \end{document}